\documentclass[onecolumn,showpacs,preprintnumbers,prc,floatfix]{revtex4}
\usepackage{amsmath,amssymb}
\usepackage{graphicx}
\usepackage{dcolumn}
\usepackage[vcentermath]{youngtab}
\usepackage[dvips]{epsfig,color}
\usepackage{bm}

\begin{document}

\title{Diquark correlations in baryons:$\;$  the Interacting Quark Diquark Model \footnote{Plenary Talk presented at The 10th International Workshop on the Physics of Excited Nucleons (NSTAR2015), May 25-28 2015, Osaka (Japan).}}   
\author{E. Santopinto}
\affiliation{INFN, Sezione di Genova, via Dodecaneso 33, 16146 Genova (Italy)}
\begin{abstract}
A review of the underlying ideas of the Interacting Quark Diquark Model (IQDM) that asses the  baryon spectroscopy  in terms of quark diquark  degrees of freedom is  given, together with a discussion of the missing resonances problem. Some ideas about its generalization the heavy baryon spectroscopy is given.s of freedom is  given, together with a discussion of the missing resonances problem. Some ideas about its generalization the heavy baryon spectroscopy is given.The results are compared to the existing experimental data. 
\end{abstract}
\maketitle

\section{Introduction}
The  notion of diquark is as old as the quark model itself. Gell-Mann \cite{gell} mentioned the possibility of diquarks in his original paper on quarks, just as the possibility of tetra and pentaquark. Soon afterwards, Ida and Kobayashi \cite{ida} and Lichtenberg and Tassie \cite{lich} introduced effective degrees of freedom of diquarks in order to describe baryons as composed of a constituent diquark and quark. Since its introduction, many articles have been written on this subject \cite{ans,Jakob:1997,Brodsky:2002,Gamberg:2003,Jaffe:2003,Wilczek,Jaffe:2004ph,Santopinto:2004,Selem:2006nd,DeGrand:2007vu,Forkel:2008un,Anisovich:2010wx}  up to the most recent ones \cite{Ferretti:2011,Santopinto:2015,desanctis}.
 But most important,
different phenomenological indications for diquark correlations have been collected during the years, such as some
regularities in hadron spectroscopy, the $\Delta I = \frac{1}{2}$  rule in weak  nonleptonic decays \cite{Neubert} , some regularities in parton distribution functions
 and in spin-dependent structure functions \cite{Close} and in the $\Lambda(1116)$ and  $\Lambda(1520)$  fragmentation functions..  Finally, although the phenomenon of color superconductivity \cite{bailingwilczek} in quark dense matter cannot be considered an argument in support of diquarks in the vacuum, it is nevertheless of interest since it stresses the important role of Cooper pairs of color superconductivity, which are color antitriplet, flavor antisymmetric, scalar diquarks.
The introduction of diquarks in hadronic physics has some similarities to that of correlated pairs in condensed matter physics (superconductivity \cite{bcs}) and in nuclear physics (interacting boson model \cite{ibm}), where effective bosons emerge from pairs of electrons \cite{coop} and nucleons \cite{ibm2}, respectively.
Most important, any interaction that binds $\pi$ and $\rho$  mesons in the rainbow-ladder approximation of the DSE will produce diquarks as can be seen  in Ref. \cite{Roberts}, moreover there are some indication of diquark confinement  \cite{Roberts1}.

  Recently quark-diquark effective degrees of freedom have shown their usefulness also in the study of transversity problems and fragmentation functions (see  Ref. \cite{Bacchetta}), even in an oversimplified form, i.e. with the spatial part of the quark-diquark ground state wave function parametrized by means of a gaussian.  Even if the microscopic origin of that effective degree of freedom of diquark, it is not completely clear, nevertheless, as in nuclear physics, one may attempt to correlate the data in terms of a phenomenological model, and in many cases it has already shown it usefulness. In this article, we will review the Interacting Quark Diquark model in its original formulation  \cite{Santopinto:2004}, since the Point Form relativistic reformulation it has been already discussed in  another contribution of the same conference (see Ferretti and Santopinto). We shall  focus  on  its differences  and extension to the strange spectra.
  We will point out some important consequences  on the ratio of the electric and magnetic form factor of the proton, that is a  zero at $Q^2= 8\ GeV^2$, while impossible with three quark model such as  that of Ref.Ref. \cite{Santopinto:Vasallo}. The new $12 \ GeV^2$ experiment  planned at Jlab will eventually shed light on the 
  diquark structure  of the nucleon. \\
 Finally by comparing  the number of   $\Lambda$'s states predicted by the  relativistic Interacting Quark Diquark model,
  that are only a subset of  those predicted by  three-quark models, 
 we  will  try to suggest  a next generation Pentaquark analysis that  evaluates the systematic error  on the background  due to  the missing
  $\Lambda$'s states.  The missing $\Lambda$ resonances  can not change
  the structures seen in the Dalitz Plot but eventually only modify some
  parameters.

  The various aspects of the hadron structures have been investigated
by many experimental and theoretical approaches in the last years.
The observations of the hadron states with an exotic structure have attracted  a lot  interest. In particular, regarding the light flavor region, we can remind  the exotic states found in the accelerator facilities such as the scalar mesons $a_0(980)$ and $f_0(980)$,or the $\Lambda(1405)$ 
which are expected to have an exotic structure as multiquarks, hadronic molecules,  but  also  hybrid states  and so forth \cite{Klempt:2007cp,Brambilla:2010cs}, while  regarding the 
 exotic heavy hadrons, we can cite states such as the $Z_c$ \cite{Ablikim:2013mio,Liu:2013dau} and $Z^{(\prime)}_b$ \cite{Belle:2011aa} which  can not  be explained by the simple quark model picture.  
In addition, there are the  pentaquark states, seen by   LHCb~\cite{Aaij:2015tga},  with  quark content  $c\bar{c}uud$.

In parallel, recently,  theoretical approaches based on QCD have been  strongly developed.
The chiral effective field theory respecting  the chiral symmetry
provides the hadron-hadron scatterings at  low energy
with the Nambu-Goldstone bosons exchange. This  is a powerful tool to investigate hadronic molecules as the meson-meson~\cite{Oller:1997ti,Wang:2013kva,Baru:2015nea}, meson-baryon~\cite{Hyodo:2011ur,Yamaguchi:2011xb}, and baryon-baryon~\cite{Machleidt:2011zz,Haidenbauer:2011za} states appearing near  thresholds.  On the other side, Lattice QCD performs ab initio calculation for hadron spectroscopy, even if  it is not easy to approach hadron states at the physical pion mass or with heavy flavor. Nevertheless, the  recent progress of the Lattice simulations  are really impressive and
hadron structures and interactions have been discussed  extensively  in  Refs.~\cite{Dudek:2009qf,Aoki:2012oma}.\\
 Finally , in the last part of this article, we will discuss  briefly some new results obtained within the  formalism of the Unquenched Quark Model (UQM):  when  LQCD or Chiral effective models can not be applied,  it   can provide  anyway predictions,  making up  with the three quark model defects.\\
 \section{The Interacting quark diquark model}
      
The  model will be discussed as an attempt to arrive to a systematic description  and correlation of data in term of q-diquark effective degrees of freedom
 by formulating a quark- diquark model with explicit interactions, in particular with a direct and an exchange interaction. We will  show the spectrum which emerges from this model.  In respect to the  prediction shown in Ref. \cite{Santopinto:2004} we have  extended our calculation up to 2.4 GeV, and so we have predicted more states.

Up to an energy of  about $2 GeV$, the diquark can be described as two correlated quarks with no internal spatial excitations \cite{Santopinto:2004,Ferretti:2011}. 
Then, its color-spin-flavor wave function must be antisymmetric. 
Moreover, as we consider only light baryons, made up of $u$, $d$, $s$ quarks, the internal group is restricted to SU$_{\mbox{sf}}$(6). 
If we denote spin by its value, flavor and color by the dimension of the representation, the quark has spin $s_2 = \frac{1}{2}$, flavor $F_2={\bf {3}}$, and color $C_2 = {\bf {3}}$. 
The diquark must transform as ${\bf {\overline{3}}}$ under SU$_{\mbox{c}}$(3), hadrons being color singlets. Then, one only has the symmetric SU$_{\mbox{sf}}$(6) representation $\mbox{{\boldmath{$21$}}}_{\mbox{sf}}$(S), containing $s_1=0$, $F_1={\bf {\overline{3}}}$, and $s_1=1$, $F_1={\bf {6}}$, i.e. the scalar and axial-vector diquarks, respectively. This is because we think of the diquark as two correlated quarks in an antisymmetric nonexcited state.
We assume that the baryons are composed of a valence quark and a valence diquark,

The relative configurations of two body can be described by the relative coordinate $\vec{r}$ and its conjugate momenta $\vec{p}$.
The Hamiltonian  contains a direct and an exchange interaction. The direct interaction is  coulomb plus linear interaction, while the exchange one is of the type spin-spin, isospin-isospin etc.  A contact term has to be present to describe the splitting between the nucleon and the $\Delta$:
\begin{eqnarray}
H =E_{0}+\frac{p^{2}}{2m}-\frac{\tau }{r}+{\beta }r+(B+C\delta
_{0})\delta _{S_{12},1}~~~~~~~~~  \nonumber \\
+(-1)^{l+1}2Ae^{-\alpha r}[\vec{s_{12}}\cdot \vec{s_{3}}+\vec{t_{12}}%
\cdot \vec{t_{3}}+2\vec{s_{12}}\cdot \vec{s_{3}}~\vec{t_{12}}\cdot \vec{t_{3}%
},
\end{eqnarray}
For a purely Coulomb-like interaction the problem is
analytically solvable. The solution is trivial, with eigenvalues 
\begin{equation}
E_{n,l}~=~-\frac{\tau ^{2}m}{2~n^{2}}~~~~,~n~=~1,~2~...~~~~~.
\end{equation}
Here $m$ is the reduced mass of the diquark-quark configuration and $n$ the
principal quantum number. The eigenfunctions are the   usual  Coulomb functions 
\begin{equation}
R_{n,l}(r)=\sqrt{\frac{(n-l-1)!(2g)^{3}}{2n[(n+l)!]^{3}}}(2gr)^{l}~e^{-gr}L_{n-l-1}^{2l+1}(2gr),
\end{equation}
where for the associated Laguerre polynomials we use the notation  of Ref. {morse} and $g=\frac{\tau m}{n}$.
We treat all the other interactions as perturbations, so the  model is completely analytical.
The matrix elements of $\beta r$ can be evaluated  in closed form as 
\begin{equation}
\Delta E_{n,l}=\int_{0}^{\infty }\beta r[R_{n,l}(r)]^{2}r^2dr=\frac{%
\beta }{2m\tau }[3n^{2}-l(l+1)].
\end{equation}

Next comes the exchange interaction of Eq. (5). The spin-isospin part
is obviously diagonal in the basis of Eq. (7)
\begin{eqnarray}
\langle \vec{s}_{12}\cdot \vec{s}_{3}\rangle =\frac{1}{2}\left[
S(S+1)-s_{12}(s_{12}+1)-s_{3}(s_{3}+1)\right]  \nonumber \\
\langle \vec{t}_{12}\cdot \vec{t}_{3}\rangle =\frac{1}{2}\left[
T(T+1)-t_{12}(t_{12}+1)-t_{3}(t_{3}+1)\right] .
\end{eqnarray}
To complete the evaluation, we need the matrix elements of the exponential.
These can be obtained in analytic form 
\begin{equation}
I_{n,l}(\alpha )~=~\int_{0}^{\infty }~e^{-\alpha ~r}~[R_{n,l}(r)]^{2}r^2dr~~.
\end{equation}
The results are straightforward. Here, by way of example, we quote the result
for $l =n-1$ 
\begin{equation}
I_{n,l=n-1}(\alpha )~=~(\frac{1}{1+\frac{n~\alpha }{2\tau ~m}})^{2n+1}~~~.
\end{equation}
Our results are in present in Tables \ref{tab:nuc-spectrum} and \ref{tab:del-spectrum}.

\begin{table}
\centering 
\vspace{15pt} 
\begin{tabular}{ccccc}
\hline
\hline
& & & &  \\
Baryon $L_{2I,2J}$ & Status & Mass & $J^p$ & $M_{{\rm cal}}$ \\
& & (MeV) & & (MeV)  \\
\hline
& & & &  \\
$N( 939)P_{11}$   & **** & 939       & $1/2^+$ &  940 \\
$N(1440)P_{11}$   & **** & 1410-1450 & $1/2^+$ & 1538 \\
$N(1520)D_{13}$   & **** & 1510-1520 & $3/2^-$ & 1543 \\
$N(1535)S_{11}$   & **** & 1525-1545 & $1/2^-$ & 1538 \\
$N(1650)S_{11}$   & **** & 1645-1670 & $1/2^-$ & 1673 \\
$N(1675)D_{15}$   & **** & 1670-1680 & $5/2^-$ & 1673 \\
$N(1680)F_{15}$   & **** & 1680-1690 & $5/2^+$ & 1675 \\
$N(1700)D_{13}$   &  *** & 1650-1750 & $3/2^-$ & 1673 \\
$N(1710)P_{11}$   &  *** & 1680-1740 & $1/2^+$ & 1640 \\
$N(1720)P_{13}$   & **** & 1700-1750 & $3/2^+$ & 1675 \\
$N(1860)F_{15}$   & **   & 1820-1960 & $5/2^+$ & 1975 \\
$N(1875)D_{13}$   & ***  & 1820-1920 & $3/2^-$ & 1838 \\
$N(1880)P_{11}$   & **   & 1835-1905 & $1/2^+$ & 1838 \\
$N(1895)S_{11}$   & **   & 1880-1910 & $1/2^-$ & 1838 \\
$N(1900)P_{13}$   & ***  & 1875-1935 & $3/2^+$ & 1967 \\
$N(1990)F_{17}$   & **   & 1995-2125 & $7/2^+$ & 2015 \\
$N(2000)F_{15}$   & **   & 1950-2150 & $5/2^+$ & 2015 \\
$N(2040)P_{13}$   & *    & 2031-2065 & $3/2^+$ & 2015 \\
$N(2060)D_{15}$   & **   & 2045-2075 & $5/2^-$ & 2078 \\
$N(2100)P_{11}$   & **   & 2050-2200 & $1/2^+$ & 2015\\
$N(2120)D_{13}$   & **   & 2090-2210 & $3/2^-$ & 2069 \\
& & & &  \\
\hline
\end{tabular}
\caption{Mass spectrum of $N$-type resonances (up to $ 2.1$ GeV) in the interacting quark diquark model \cite{Santopinto:2004}. The value of the parameters  are those obtained and reported in Ref.[10] based on the fit of the 3 and 4  star  resonances known at the time. The table reports also the prediction for the remaining resonances, including the recent upgraded 3* $P13(1900) $. The experimental values are taken from Ref. \cite{Olive:2014}.}

\label{tab:nuc-spectrum}
\end{table}

\begin{table}
\centering 
\begin{tabular}{ccccc}
\hline
\hline
& & & &  \\
Baryon $L_{2I,2J}$ & Status & Mass & State & $M_{{\rm cal}}$ \\
 
& & (MeV) & & (MeV)  \\

\hline
& & & &  \\
$\Delta(1232)P_{33}$ & **** & 1230-1234 & $3/2^+$ & 1235 \\
$\Delta(1600)P_{33}$ &  *** & 1500-1700 & $3/2^+$ & 1709 \\
$\Delta(1620)S_{31}$ & **** & 1600-1660 & $1/2^-$ & 1673 \\
$\Delta(1700)D_{33}$ & **** & 1670-1750 & $3/2^-$ & 1673 \\
$\Delta(1900)S_{31}$ & **   & 1840-1920 & $1/2^-$ & 2003 \\
$\Delta(1905)F_{35}$ & **** & 1855-1910 & $5/2^+$ & 1930 \\ 
$\Delta(1910)P_{31}$ & **** & 1860-1910 & $1/2^+$ & 1967 \\ 
$\Delta(1920)P_{33}$ &  *** & 1900-1970 & $3/2^+$ & 1930 \\ 
$\Delta(1930)D_{35}$ &  *** & 1900-2000 & $5/2^-$ & 2003 \\ 
$\Delta(1940)D_{33}$ & **   & 1940-2060 & $3/2^-$ & 2003 \\
$\Delta(1950)F_{37}$ & **** & 1915-1950 & $7/2^+$ & 1930 \\ 
$\Delta(2000)F_{35}$ & **   & $\approx$ 2000 & $5/2^+$ &2015  \\ 
& & & &  \\
\hline
\hline
\end{tabular}
\caption{As Table \ref{tab:nuc-spectrum}, but for $\Delta$-type resonances.}
\label{tab:del-spectrum}
\end{table}

\section{The relativistic Interacting quark diquark model}

The exstention of the Interacting quark diquark model  \cite{Santopinto:2004} in Point Form can be easily done \cite{Ferretti:2011,Santopinto:2015}.
This is a potential model, constructed within the point form formalism \cite{Klink:1998zz}, where baryon resonances are described as two-body quark-diquark bound states; thus, the relative motion between the two constituents and the Hamiltonian of the model are functions of the relative coordinate $\vec r$ and its conjugate momentum $\vec q$. 
The Hamiltonian contains  just as in the 2005 paper \cite{Santopinto:2004}, the two  basic ingredients:  a Coulomb-like  plus linear confining interaction and an exchange one, depending on the spin and isospin of the quark and the diquark. 
The mass operator  is given by
\begin{equation}
	\begin{array}{rcl}
	M & = & E_0 + \sqrt{\vec q\hspace{0.08cm}^2 + m_1^2} + \sqrt{\vec q\hspace{0.08cm}^2 + m_2^2} 
	+ M_{\mbox{dir}}(r)  + M_{\mbox{ex}}(r)  
	\end{array}  \mbox{ },
	\label{eqn:H0}
\end{equation}
where $E_0$ is a constant, $M_{\mbox{dir}}(r)$ and $M_{\mbox{ex}}(r)$ the direct and the exchange diquark-quark interaction, respectively, $m_1$ and $m_2$ stand for diquark and quark masses.
The direct term we consider, 
\begin{equation}
  \label{eq:Vdir}
  M_{\mbox{dir}}(r)=-\frac{\tau}{r} \left(1 - e^{-\mu r}\right)+ \beta r ~~,
\end{equation}
is the sum of a Coulomb-like interaction with a cut off plus a linear confinement term. 
We also have an exchange interaction, since this is the crucial ingredient of a quark-diquark description of baryons that has to be extended to contain flavor   $\lambda$ matrices in such a way to be able to describe in a simultaneous way  both the non strange and the strange sector \cite{Santopinto:2004, Santopinto:2015}. 
We have also generalized  the exchange interaction in such a way to be able to describe  strange baryons, simply considering 
\begin{equation}
	\begin{array}{rcl}
	M_{\mbox{ex}}(r) & = & \left(-1 \right)^{L + 1} \mbox{ } e^{-\sigma r} \left[ A_S \mbox{ } \vec{s}_1 
	\cdot \vec{s}_2  + A_F \mbox{ } \vec{\lambda}_1^f \cdot \vec{\lambda}_2^f \mbox{ } 
	+ A_I \mbox{ } \vec{t}_1 \cdot \vec{t}_2  \right]  
	\end{array}  \mbox{ },
	\label{eqn:Vexch-strange}
\end{equation}
where $\vec{\lambda}^f$ are the SU$_{\mbox{f}}$(3) Gell-Mann matrices. 
In a certain sense, we can consider it as a G\"ursey-Radicati inspired interaction \cite{desanctis,Gursey:1992dc}.
In the nonstrange sector, we also have to keep a contact interaction  \cite{Ferretti:2011} in the mass operator
\begin{equation}
	\begin{array}{rcl}
	\label{eqn:Vcont}	
	M_{\mbox{cont}} & = & \left(\frac{m_1 m_2}{E_1 E_2}\right)^{1/2+\epsilon} \frac{\eta^3 D}{\pi^{3/2}} 
	e^{-\eta^2 r^2} \mbox{ } \delta_{L,0} \delta_{s_1,1}  \left(\frac{m_1 m_2}{E_1 E_2}\right)^{1/2+\epsilon}
	\end{array}  \mbox{ },
\end{equation}
as necessary to reproduce the $\Delta-N$ mass splitting. 

\begin{table}
	\begin{tabular}{cccccccccccccc}
		\hline
		\hline \\
		Resonance & Status & $M^{\mbox{exp.}}$ & $J^P$ & $L^P$ & $S$ & $s_1$ & $Q^2q$ & ${\bf F}$ & ${\bf {F_1}}$ & $I$ & $t_1$ & $n_r$ & $M^{\mbox{calc.}}$ \\
		&  & (MeV) &  &  &  &  &  &  & & & & & (MeV) \\ \\
		\hline \\
		$\Lambda(1116)$ $P_{01}$ & **** & 1116        & $\frac{1}{2}^+$ & $0^+$ & $\frac{1}{2}$ & 0   & $[n,n]s$   & ${\bf 8}$ & ${\bf {\bar 3}}$ & 0 & 0             & 0 & 1116  \\
		$\Lambda(1600)$ $P_{01}$ & ***  & 1560 - 1700 & $\frac{1}{2}^+$ & $0^+$ & $\frac{1}{2}$ & 0   & $[n,s]n$   & ${\bf 8}$ & ${\bf {\bar 3}}$ & 0 & $\frac{1}{2}$ & 0 & 1518  \\
		$\Lambda(1670)$ $S_{01}$ & **** & 1660 - 1680 & $\frac{1}{2}^-$ & $1^-$ & $\frac{1}{2}$ & 0   & $[n,n]s$   & ${\bf 8}$ & ${\bf {\bar 3}}$ & 0 & 0             & 0 & 1650  \\
		$\Lambda(1690)$ $D_{03}$ & **** & 1685 - 1695 & $\frac{3}{2}^-$ & $1^-$ & $\frac{1}{2}$ & 0   & $[n,n]s$   & ${\bf 8}$ & ${\bf {\bar 3}}$ & 0 & 0             & 0 & 1650  \\
		$\Lambda(1800)$ $S_{01}$ & ***  & 1720 - 1850 & $\frac{1}{2}^-$ & $1^-$ & $\frac{1}{2}$ & 0   & $[n,s]n$   & ${\bf 8}$ & ${\bf {\bar 3}}$ & 0 & $\frac{1}{2}$ & 0 & 1732  \\
		$\Lambda(1810)$ $P_{01}$ & ***  & 1750 - 1850 & $\frac{1}{2}^+$ & $0^+$ & $\frac{1}{2}$ & 0   & $[n,n]s$   & ${\bf 8}$ & ${\bf {\bar 3}}$ & 0 & 0             & 1 & 1666  \\
		$\Lambda(1820)$ $F_{05}$ & **** & 1815 - 1825 & $\frac{5}{2}^+$ & $2^+$ & $\frac{1}{2}$ & 0   & $[n,n]s$   & ${\bf 8}$ & ${\bf {\bar 3}}$ & 0 & 0             & 0 & 1896  \\
		$\Lambda(1830)$ $D_{05}$ & **** & 1810 - 1830 & $\frac{5}{2}^-$ & $1^-$ & $\frac{3}{2}$ & 1   & $\{n,s\}n$ & ${\bf 8}$ & ${\bf {6}}$      & 0 & $\frac{1}{2}$ & 0 & 1785  \\
		$\Lambda(1890)$ $P_{03}$ & **** & 1850 - 1910 & $\frac{3}{2}^+$ & $0^+$ & $\frac{3}{2}$ & 1   & $\{n,s\}n$ & ${\bf 8}$ & ${\bf {6}}$      & 0 & $\frac{1}{2}$ & 0 & 1896  \\  \\
		missing                 & --   & --          & $\frac{3}{2}^-$ & $1^-$ & $\frac{1}{2}$ & 0   & $[n,s]n$   & ${\bf 8}$ & ${\bf {\bar 3}}$ & 0 & $\frac{1}{2}$ & 0 & 1732  \\
		missing                 & --   & --          & $\frac{1}{2}^-$ & $1^-$ & $\frac{3}{2}$ & 1   & $\{n,s\}n$ & ${\bf 8}$ & ${\bf {6}}$      & 0 & $\frac{1}{2}$ & 0 & 1785  \\
		missing                 & --   & --          & $\frac{3}{2}^-$ & $1^-$ & $\frac{1}{2}$ & 0   & $[n,n]s$   & ${\bf 8}$ & ${\bf {\bar 3}}$ & 0 & 0             & 1 & 1785  \\
		missing                 & --   & --          & $\frac{1}{2}^+$ & $0^+$ & $\frac{1}{2}$ & 1   & $\{n,s\}n$ & ${\bf 8}$ & ${\bf {6}}$      & 0 & $\frac{1}{2}$ & 0 & 1955  \\ 
		missing                 & --   & --          & $\frac{1}{2}^+$ & $0^+$ & $\frac{1}{2}$ & 0   & $[n,s]n$   & ${\bf 8}$ & ${\bf {\bar 3}}$ & 0 & $\frac{1}{2}$ & 1 & 1960  \\
		missing                 & --   & --          & $\frac{1}{2}^-$ & $1^-$ & $\frac{1}{2}$ & 1   & $\{n,s\}n$ & ${\bf 8}$ & ${\bf {6}}$      & 0 & $\frac{1}{2}$ & 0 & 1969  \\
		missing                 & --   & --          & $\frac{3}{2}^-$ & $1^-$ & $\frac{1}{2}$ & 1   & $\{n,s\}n$ & ${\bf 8}$ & ${\bf {6}}$      & 0 & $\frac{1}{2}$ & 0 & 1969  \\  \\
		$\Lambda^*(1405)$ $S_{01}$ & **** & 1402 - 1410 & $\frac{1}{2}^-$ & $1^-$ & $\frac{1}{2}$ & 0 & $[n,n]s$   & ${\bf 1}$ & ${\bf {\bar 3}}$ & 0 & 0             & 0 & 1431  \\
		$\Lambda^*(1520)$ $D_{03}$ & **** & 1519 - 1521 & $\frac{3}{2}^-$ & $1^-$ & $\frac{1}{2}$ & 0 & $[n,n]s$   & ${\bf 1}$ & ${\bf {\bar 3}}$ & 0 & 0             & 0 & 1431
		
		\\ 
		
 missing                 & --     & --          & $\frac{1}{2}^-$ & $1^-$ & $\frac{1}{2}$ & 0 & $[n,s]n$   & ${\bf 1}$ & ${\bf {\bar 3}}$ & 0 & $\frac{1}{2}$ & 0 & 1443  \\
 missing                 & --     & --          & $\frac{3}{2}^-$ & $1^-$ & $\frac{1}{2}$ & 0 & $[n,s]n$   & ${\bf 1}$ & ${\bf {\bar 3}}$ & 0 & $\frac{1}{2}$ & 0 & 1443  \\
  missing                 & --     & --         & $\frac{1}{2}^-$ & $1^-$ & $\frac{1}{2}$ & 0 & $[n,n]s$  & ${\bf 1}$ & ${\bf {\bar 3}}$ & 0 & 0               & 1 & 1854  \\
	  missing                 & --     & --         & $\frac{3}{2}^-$ & $1^-$ & $\frac{1}{2}$ & 0 & $[n,n]s$  & ${\bf 1}$ & ${\bf {\bar 3}}$ & 0 & 0               & 1 & 1854  \\
  missing                 & --     & --         & $\frac{1}{2}^-$ & $1^-$ & $\frac{1}{2}$ & 0 & $[n,s]n$  & ${\bf 1}$ & ${\bf {\bar 3}}$ & 0 & $\frac{1}{2}$   & 1 & 1928  \\ 
  missing                 & --     & --         & $\frac{3}{2}^-$ & $1^-$ & $\frac{1}{2}$ & 0 & $[n,s]n$  & ${\bf 1}$ & ${\bf {\bar 3}}$ & 0 & $\frac{1}{2}$   & 1 & 1928  \\  \\

& & & &  \\
\hline
\hline
\end{tabular}
\caption{Mass predictions \cite{Santopinto:2015}   for $\Lambda$-type resonances compared with PDG data; APS copyright.}
\label{tab:lam-spectrum}
\end{table}

The results  for the strange and non-strange baryon spectra from Ref. \cite{Santopinto:2015, Santopinto:2004} (See Tables \ref{tab:nuc-spectrum},\ref{tab:del-spectrum} and \ref{tab:lam-spectrum}) were obtained by diagonalizing the mass operator of Eq. (\ref{eqn:H0}) by means of a numerical variational procedure, based on harmonic oscillator trial wave functions. With a basis of 150 harmonic oscillator shells, the results converge very well.

 It is clear that a larger number of experiments and analysis, looking for missing resonances, are necessary because many aspects of hadron spectroscopy are still unclear. In particular the number of 
$\Lambda $ states   reported by the PDG are still very few in respect to the predictions both of the Lattice QCD and by the models. In particular the  
relativistic version of the  interacting quark diquark model  predict  seven $\Lambda$  missing  states belonging to the octet and other  six  missing states  belonging to the singlet (considering only states under  2.0 GeV), otherwise much more states should be considered. Typical three quark models  will predict much more $\Lambda $  missing states. New experiments should be dedicated to the hunting of those elusive missing $\Lambda$ states.
Incidentally, we observe that in the $LHC_b$ pentaquark finding \cite{LHC_b}, the authors have included in the fit, correctly, only the  experimentally well established  $\Lambda$ states,  that means that   the one * star states have been excluded.  We  observe  that the eventual  future discover of extra 
$\Lambda$ states or even the confirmation of the one * star states , even  not  able, for sure,  to modify the structures present in the Dalitz plot, can eventually change  the parameters of the two Pentaquark states.


It is also worthwhile noting that in our model  \cite{Santopinto:2015}  $\Lambda(1116)$ and $\Lambda^*(1520)$ are described as bound states of a scalar diquark $[n,n]$ and a quark $s$, where the quark-diquark system is in $S$ or $P$-wave, respectively \cite{Santopinto:2015} . This is in accordance with the observations of Refs. \cite{Jaffe:2004ph,Selem:2006nd} on $\Lambda$'s fragmentation functions, that the two resonances can be described as $[n,n]-s$ systems.

\begin{figure}[h]
\begin{center}
\includegraphics[width=7cm]{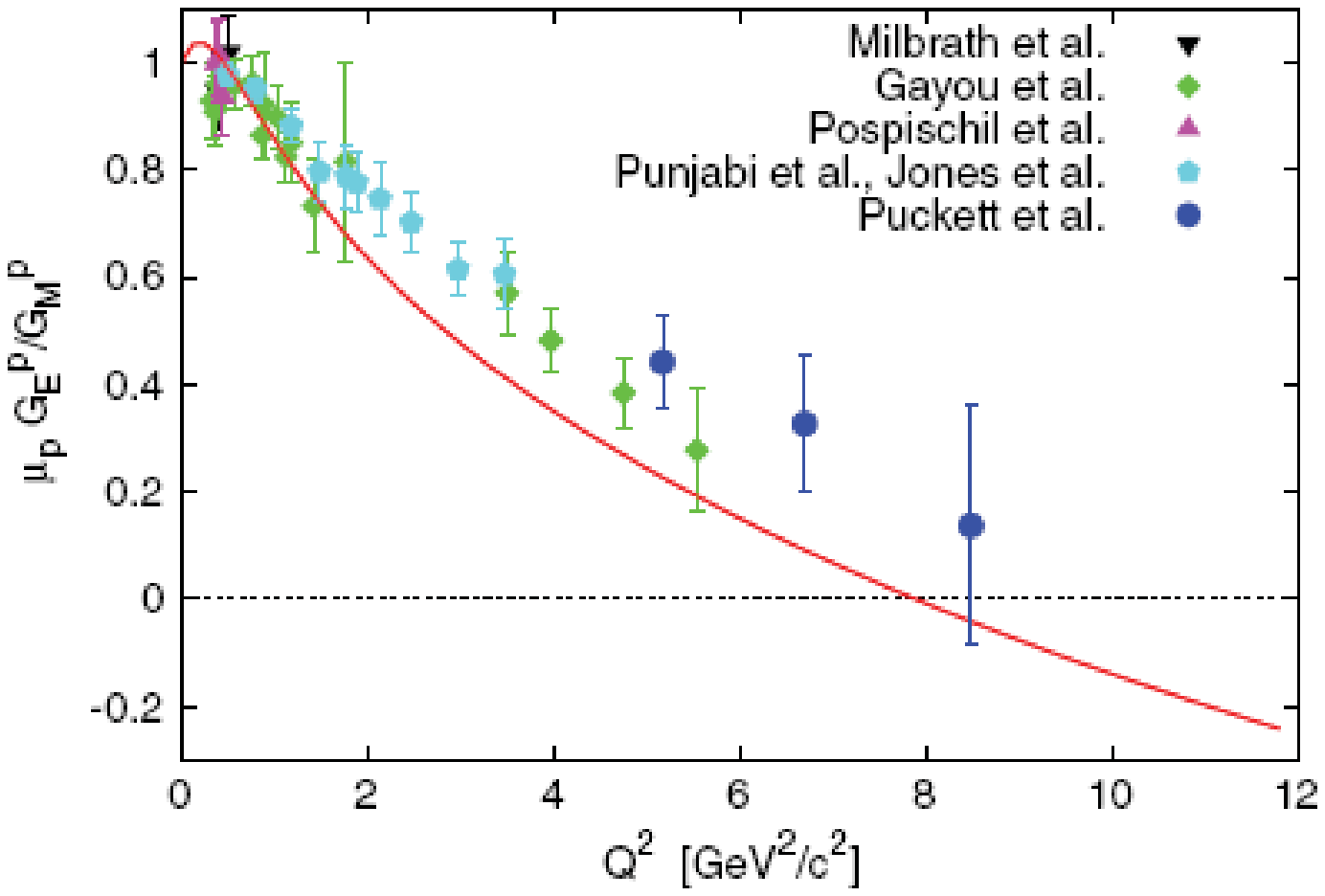}
\end{center}
\caption{\label{ratiomp1}Ratio $\mu_pG^p_E(Q^2)/G^p_M(Q^2)$, the solid line correspond to the relativistic quark-diquark calculation,
figure taken from Ref. \cite{Ferretti_ff:2011}; APS copyright.}
\begin{center}
\includegraphics[width=7cm]{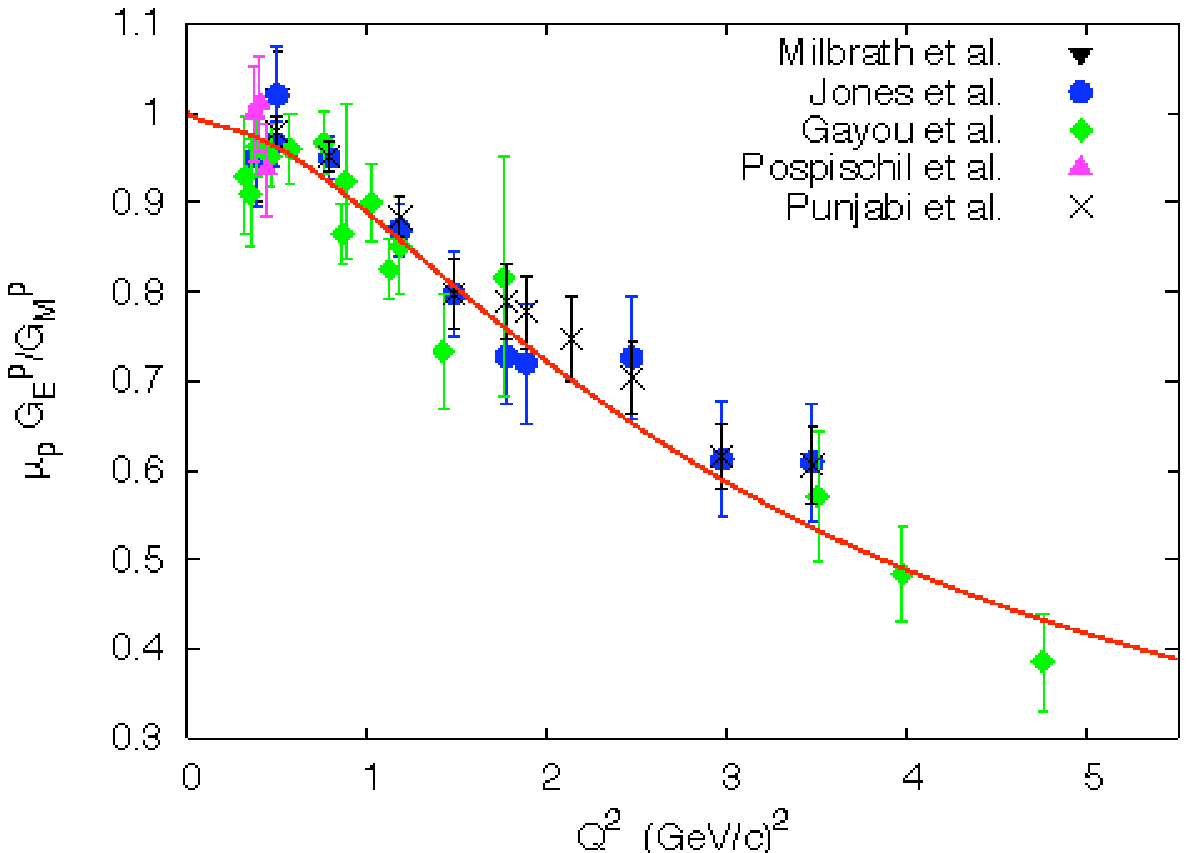}
\end{center}
\caption{\label{ratiomp2} Ratio $\mu_pG^p_E(Q^2)/G^p_M(Q^2)$, the solid line correspond to the relativistic Hypercetral quark model, Figure taken from 
Ref. \cite{Santopinto:Vasallo};  APS copyright.}
\end{figure}

We should also underline that the interacting quark-diquark model  gives origin to wave functions that can describe in a  reasonable way the elastic electromagnetic form factors of the nucleon . In particular they give origin to  a reproduction of the existing data for the ratio of the electric and magnetic form factor of the proton that predict a zero at  $Q^2= 8\  GeV^2$(see Fig. \ref{ratiomp1})  like in vector meson parametrizations. On the contrary,  we have found impossible to get this zero with a three quark model  (see Fig. \ref{ratiomp2}) . The new experiment planned at JLab will be able to distinguish  between the two scenarios ruling out one of the two models.  

\section{The Unquenched  Quark Model}
The behavior of observables such as the spectrum and the magnetic moments of hadrons are well reproduced by   the   constituent quark model (CQM) \cite{Eichten:1974af,Isgur:1979be,Godfrey:1985xj,Capstick:1986bm,Giannini:2001kb,Glozman-Riska,Loring:2001kx,Ferretti:2011,Galata:2012xt,BIL}, but it neglects quark-antiquark pair-creation (or continuum-coupling) effects.
The unquenching of the quark model for hadrons is a way to take these components into account.

The unquenching of  CQM were initially   done by T\"ornqvist and collaborators, who used  an unitarized quark model \cite{Ono:1983rd,Tornqvist}, while 
Van Beveren and Rupp used an 
t-matrix approach \cite{vanBeveren:1979bd,vanBeveren:1986ea}.
These techniques were applied to study of scalar meson nonet ($a_0$, $f_0$, etc.) of Ref. \cite{vanBeveren:1986ea,Tornqvist:1995kr} in which the loop contributions are given by the hadronic intermediate states that each meson can access. It is via these hadronic loops that the bare states become ``dressed'' and  the hadronic loop contributions totally dominate the dynamics of the process.  A similar approach was developed by Pennington in Ref. \cite{Pennington:2002}, where they have  investigated the dynamical generation of the scalar mesons by initially inserting only one ``bare seed''.  Also, the strangeness content of the nucleon and electromagnetic form factors were  investigated in  \cite{Bijker:2012zza}, whereas  Capstick and Morel in Ref. \cite{Capstick} analyzed  
baryon meson loop effects on the spectrum of nonstrange baryons.   
In the meson sector, Eichten {\it et al.} explored the influence of the open-charm channels on the charmonium properties using the Cornell coupled-channel model \cite{Eichten:1974af} to assess departures from the single-channel  potential-model expectations.

In this work we present  the latest applications of the UQM to study the orbital angular momenta contribution to the spin of the proton 
 in which the effects of the sea quarks were introduced  into the CQM in a systematic way and the wave functions  given explicitly. In another contribution of the same workshop ( see Hugo Garcia et al.)  are on the contrary discussed the flavor asymmetry  and strangeness  of the proton.
  Finally, the UQM is applied  to describe meson observables and the spectroscopy of the charmonium  and bottomonium, developing the formalism to take into account in a systematic way,  the continuum components.  	

\section{The UQM formalism  }
\label{Sec:formalism}
In the UQM for baryons \cite{Bijker:2012zza,Santopinto:2010zza,Bijker:2009up,Bijker:210} and mesons \cite{bottomonium,charmonium,Ferretti:2013vua,Ferretti:2014xqa}, the hadron wave function is made up of a zeroth order $qqq$ ($q \bar q$) configuration plus a sum over the possible higher Fock components, due to the creation of $^{3}P_0$ $q \bar q$ pairs. Thus,  we have 
\begin{eqnarray} 
	\label{eqn:Psi-A}
	\mid \psi_A \rangle
	 ={\cal N} \left[ \mid A \rangle 
	+ \sum_{BC \ell J} \int d \vec{K} \, k^2 dk \, \mid BC \ell J;\vec{K} k \rangle \right.
	\left.  \frac{ \langle BC \ell J;\vec{K} k \mid T^{\dagger} \mid A \rangle } 
	{E_a - E_b - E_c} \right] ~, 
\end{eqnarray}
where $T^{\dagger}$ stands for the $^{3}P_0$ quark-antiquark pair-creation operator \cite{bottomonium,charmonium,Ferretti:2013vua,Ferretti:2014xqa}, $A$ is the baryon/meson, $B$ and $C$ represent the intermediate state hadrons. $E_a$, $E_b$ and $E_c$ are the corresponding energies, $k$ and $\ell$ the relative radial momentum and orbital angular momentum between $B$ and $C$ and $\vec{J} = \vec{J}_b + \vec{J}_c + \vec{\ell}$ is the total angular momentum. 
It is worthwhile noting that in Refs. \cite{bottomonium,charmonium,Ferretti:2013vua,Ferretti:2014xqa}, the constant pair-creation strength in the operator (\ref{eqn:Psi-A}) was substituted with an effective one, to suppress unphysical heavy quark pair-creation. 

 
The introduction of continuum effects in the CQM can thus be essential to study observables that only depend on $q \bar q$ sea pairs, like the strangeness content of the nucleon electromagnetic form factors \cite{Bijker:2012zza} or the flavor asymmetry of the nucleon sea \cite{Santopinto:2010zza} it has been discussed in another contribution of the same conference (see Garc\'ia-Tecocoatzi {et al.}) . 
The  continuum effects can give  important corrections to baryon/meson observables, like the self-energy corrections to meson masses \cite{bottomonium,charmonium,Ferretti:2013vua,Ferretti:2014xqa} or the importance of the orbital angular momentum in the spin of the proton \cite{Bijker:2009up}. 

\section{ Orbital Angular momenta contribution to Proton spin  in the UQM formalism}

The inclusion of the continuum higher Fock components  has a dramatic effect on the spin content of the 
proton \cite{Bijker:210}. Whereas in the CQM the proton spin is carried entirely by the (valence) quarks, 
while in the unquenched calculation 67.6 \% is carried by 
the quark and antiquark spins and the remaining 32.4 \% by orbital angular momentum. 
The orbital angular momentum due to the relative motion of the 
baryon with respect to the meson accounts for 31.7 \% of the proton spin, whereas the orbitally 
excited baryons and mesons in the intermediate state only contribute 0.7 \%. Finally we note, 
that the orbital angular momentum arises almost entirely from the relative motion of the nucleon 
and $\Delta$ resonance with respect to the $\pi$-meson in the intermediate states.

\section{Self-energy corrections in the UQM}
The formalism  was used to compute the charmonium  ($c \bar c$) and
bottomonium ($b \bar b$) spectra with self-energy corrections, due to continuum coupling effects  \cite{bottomonium,charmonium,Ferretti:2013vua,Ferretti:2014xqa}. 
In the UQM, the physical mass of a meson, 
\begin{equation}
	\label{eqn:self-trascendental}
	M_a = E_a + \Sigma(E_a)  \mbox{ },
\end{equation}
is given by the sum of two terms: a bare energy, $E_a$, calculated within a potential model \cite{Godfrey:1985xj}, and a self energy correction, 
\begin{equation}
	\label{eqn:self-a}
	\Sigma(E_a) = \sum_{BC\ell J} \int_0^{\infty} k^2 dk \mbox{ } \frac{\left| M_{A \rightarrow BC}(k) \right|^2}{E_a - E_b - E_c}  \mbox{ },
\end{equation}
computed within the UQM formalism. 

\begin{figure}[h]
\begin{center}
\includegraphics[width=16pc]{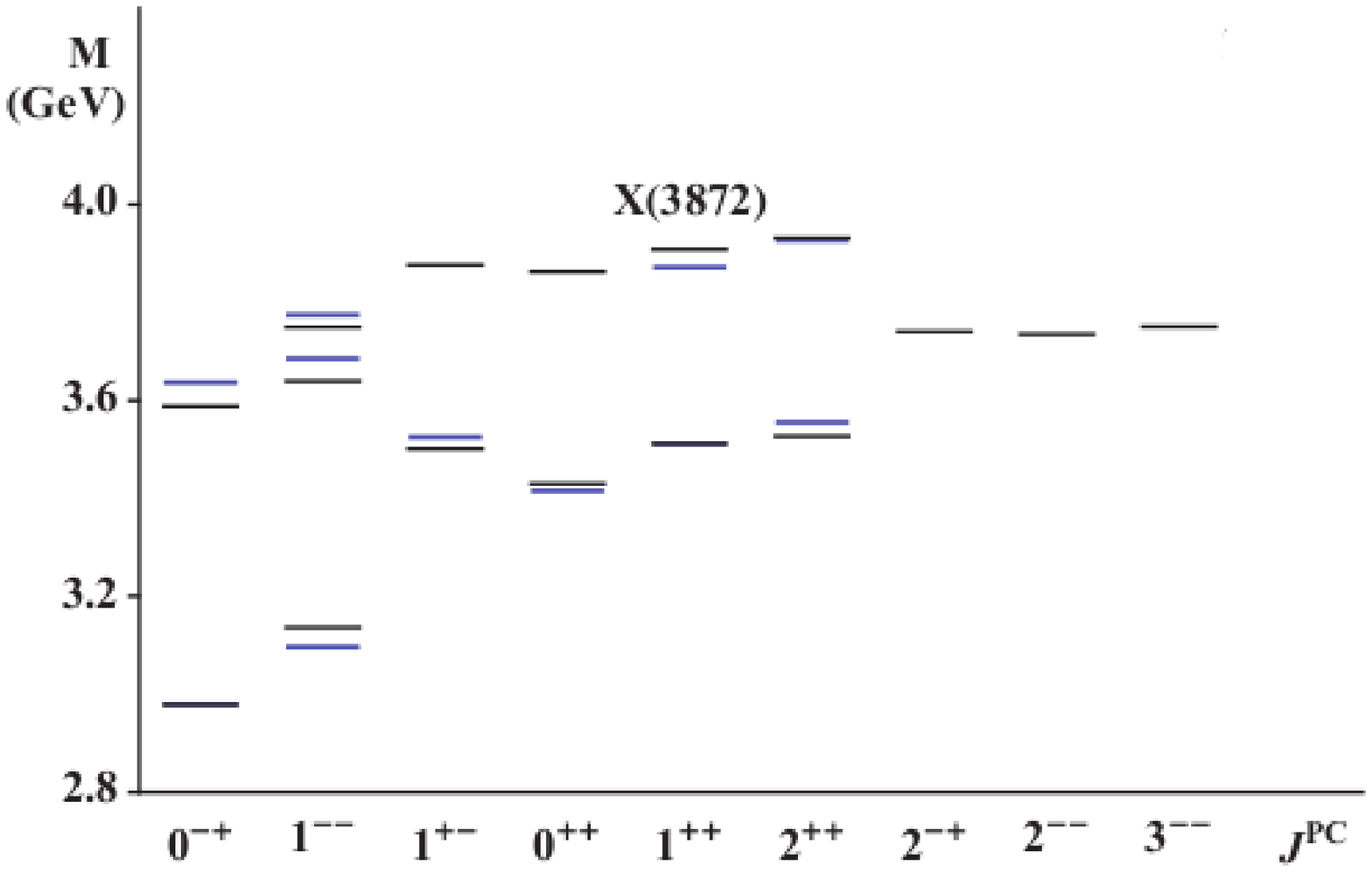}
\end{center}

 \caption{\label{charm} Charmonium spectrum with self energies corrections. 
  Black lines are theoretical predictions and blue lines are experimental data available. Figure taken from Ref. \cite{charmonium}; APS copyright.}
\begin{center}
\includegraphics[width=18pc]{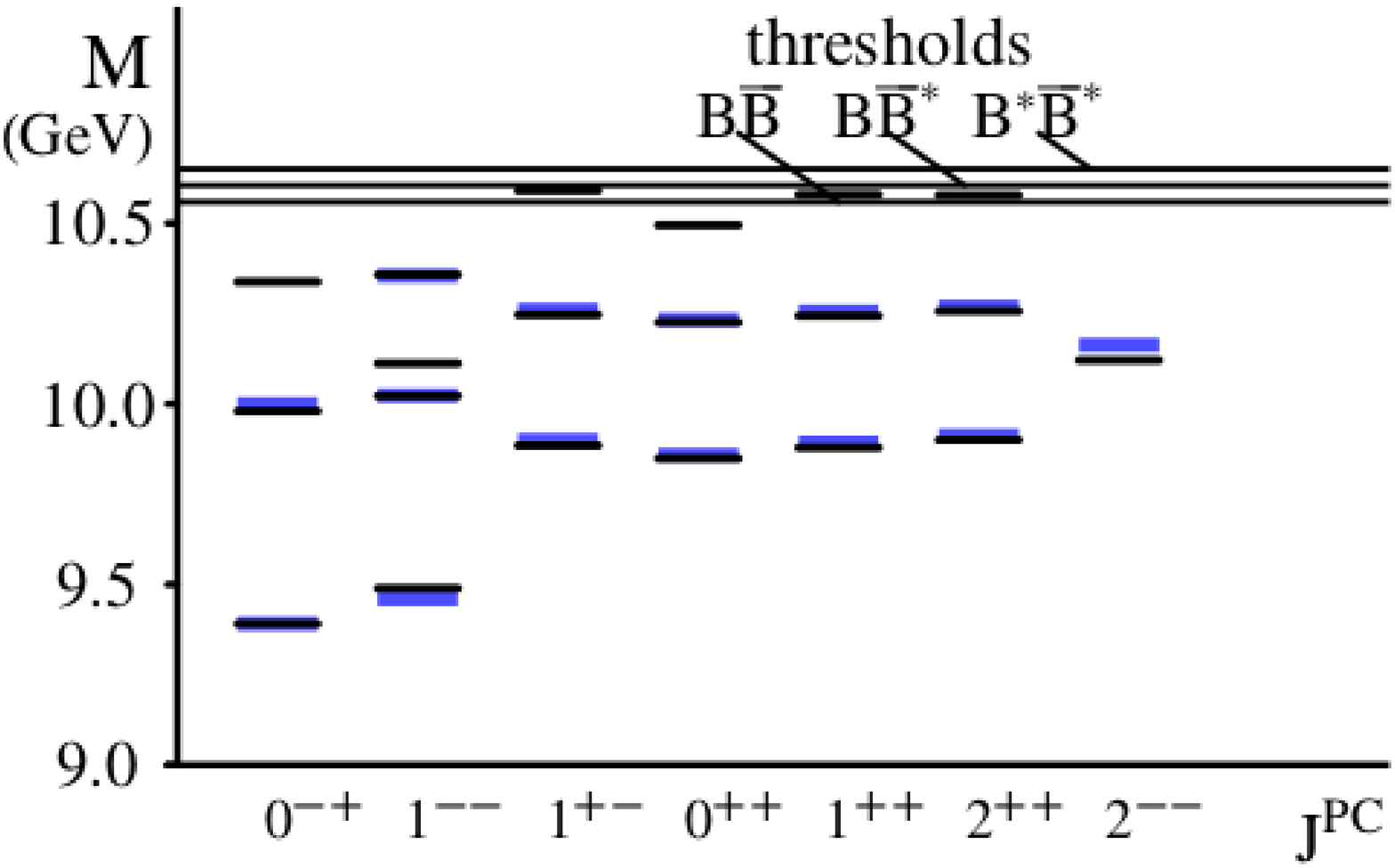}
\end{center}

 \caption{\label{botton}Bottomonium spectrum with self energies corrections.
 Black lines are theoretical predictions and blue lines are experimental data available. Figure taken from Ref. \cite{Ferretti:2013vua}; APS copyright. 
 }

\end{figure}
Our results for the self energies corrections  of charmonia \cite{charmonium,Ferretti:2014xqa} and bottomonia \cite{bottomonium,Ferretti:2013vua,Ferretti:2014xqa} spectrums, 
are shown in  figures \ref{charm} and \ref{botton}.


In our framework  the $X(3872)$ can be interpreted as a $c \bar c$ core [the $\chi_{c1}(2^3P_1)$], plus higher Fock components due to the coupling to the meson-meson continuum. In Ref. \cite{Ferretti:2014xqa}  we  were the first to predict  analogous states (as $X(3872)$) with strong continuum components  in the bottomonium sector but in the 
$\chi_{b1}(3^3P_1)$ sector, due to opening of threshold of $B\bar{B}$,  $B\bar{B}^*$ and $B^*\bar{B}^*$. We expect similar interesting effects near threshold also in the $N^*$ sector.


It is interesting to compare the present results to those of the main three-quark quark models \cite{IK,CI,HC,GR,LMP}. It is clear that a larger number of experiments and analyses, looking for missing resonances, are necessary because many aspects of hadron spectroscopy are still unclear \cite{Hugo}.

The present work can be expanded to include charmed and/or bottomed baryons \cite{FS-inprep}, which can be quite interesting in light of the recent experimental effort to study the properties of heavy hadrons. 



\end{document}